\documentclass[preprint, 3p, twocolumn]{elsarticle}

\usepackage{graphicx}
\usepackage[T1]{fontenc}
\usepackage{amsmath,amsfonts,amssymb}
\usepackage{color}
\usepackage[breaklinks,colorlinks,urlcolor=blue,citecolor=magenta,linkcolor=magenta,hypertexnames=false]{hyperref}
\usepackage{verbatim}
\usepackage{orcidlink}

\biboptions{numbers,sort&compress}

\newcommand{\be}{\begin{equation}} 
\newcommand{\ee}{\end{equation}}
\newcommand{\bea}{\begin{equation}\begin{aligned}} 
\newcommand{\eea}{\end{aligned}\end{equation}}

\newcommand{\td}{{\rm d}}

\newcommand{\papertitle}{Black holes and gravitational waves from phase transitions in realistic models
}

\begin{document}

\title{\papertitle}

\author[1]{M. Lewicki}
\ead{marek.lewicki@fuw.edu.pl}
\author[1]{P. Toczek\corref{cor1}}
\ead{piotr.toczek@fuw.edu.pl}
\author[2,3,4]{V. Vaskonen}
\ead{ville.vaskonen@pd.infn.it}
\cortext[cor1]{Corresponding author}

\affiliation[1]{organization={Faculty of Physics, University of Warsaw},
addressline={ul. Pasteura 5},
city={02-093 Warsaw},
country={Poland}}

\affiliation[2]{organization={Dipartimento di Fisica e Astronomia, Universit\`a degli Studi di Padova},
addressline={Via Marzolo 8},
city={35131 Padova},
country={Italy}}

\affiliation[3]{organization={Istituto Nazionale di Fisica Nucleare, Sezione di Padova},
addressline={Via Marzolo 8},
city={35131 Padova},
country={Italy}}

\affiliation[4]{organization={Keemilise ja bioloogilise f\"u\"usika instituut},
addressline={R\"avala pst. 10},
city={10143 Tallinn},
country={Estonia}}

\begin{abstract}
We study realistic models predicting primordial black hole (PBH) formation from density fluctuations generated in a first-order phase transition. We show that the second-order correction in the expansion of the bubble nucleation rate is necessary for accurate predictions and quantify its impact on the abundance of PBHs and gravitational waves (GWs). We find that the distribution of the fluctuations becomes more Gaussian as the second-order term increases. Consequently, models that predict the same PBH abundances can produce different GW spectra.
\end{abstract}


\maketitle

\vspace{5pt}\noindent\textbf{Introduction --} Primordial black holes (PBHs), that could have originated from various processes before the Big Bang nucleosynthesis (BBN), are perceived as an interesting candidate for dark matter (DM). Their abundance is constrained by various observations~\cite{Carr:2020gox} but they can constitute all DM in the asteroidal mass window. They may have formed in the early universe in the collapse of large density perturbations~\cite{Hawking:1971ei,Carr:1974nx}. These density perturbations could have originated from primordial inflation if, for example, the inflaton potential included an inflection point~\cite{Garcia-Bellido:1996mdl,Clesse:2015wea,Garcia-Bellido:2017mdw,Domcke:2017fix,Ezquiaga:2017fvi,Germani:2017bcs,Motohashi:2017kbs,Kannike:2017bxn,Karam:2022nym,Balaji:2022rsy,Qin:2023lgo,Ahmed:2024tlw}.

PBHs could also have formed as a consequence of an early universe first-order phase transition~\cite{Hawking:1982ga, Kodama:1982sf, Lewicki:2023ioy, Liu:2021svg, Kawana:2022olo, Gouttenoire:2023naa, Salvio:2023ynn, Baldes:2023rqv, Lewicki:2024ghw, Kanemura:2024pae, Balaji:2024rvo, Goncalves:2024vkj, Hashino:2021qoq}. Such transitions proceed by nucleation of bubbles of the true vacuum that start to expand and eventually fill the whole universe. The transition can be preceded by a period of thermal inflation if the transition is delayed so much that the vacuum energy of the false vacuum starts to dominate over the radiation energy. Then, the dominant mechanism for PBH formation is the same as in the case of primordial inflation. The source of the large fluctuations, in this case, is the Poisson nature of the bubble nucleation process~\cite{Liu:2021svg, Kawana:2022olo, Gouttenoire:2023naa, Baldes:2023rqv, Lewicki:2024ghw}. The largest fluctuations are obtained in slow transitions where each comoving Hubble patch includes a small number of bubbles.

In our previous work~\cite{Lewicki:2024ghw}, we improved the computation of the PBH abundance done in~\cite{Liu:2021svg, Kawana:2022olo, Gouttenoire:2023naa, Baldes:2023rqv} by accounting for fluctuations in the nucleation times of several bubbles and evaluating the fluctuations at all relevant scales that exit the horizon during thermal inflation. Furthermore, we showed that the gravitational wave (GW) spectrum from the transition in the regime relevant for PBH formation has a characteristic double peak structure, with the lower frequency peak arising from the large fluctuations and the higher frequency peak from bubble collisions.

In~\cite{Lewicki:2024ghw} we also developed a publicly available \texttt{C++} code \href{https://github.com/vianvask/deltaPT}{deltaPT} that allows for fast computation of the evolution of the energy densities across the phase transition. Similarly as in previous studies~\cite{Liu:2021svg, Kawana:2022olo, Gouttenoire:2023naa, Baldes:2023rqv}, we applied this computation for the case where the bubble nucleation rate can be approximated at $\Gamma\propto e^{\beta t}$. 

In this work, we go beyond the first-order approximation of the nucleation action. We show that in realistic models where the transition can be sufficiently slow and strongly supercooled, the second-order term needs to be taken into account, $\Gamma\propto e^{\beta t - \gamma^2 t^2/2}$. We compute the distribution of the density fluctuations and the resulting PBH and GW abundances with this approximation of the nucleation rate. On the model side, we treat consistently the impact of bubble nucleation on the expansion rate throughout the transition and find that the standard percolation criterion~\cite{Turner:1992tz,Ellis:2018mja} coincides with the mean bubble separation equal to the Hubble horizon size $RH=1$.

\vspace{5pt}\noindent\textbf{Formation of inhomogeneities --} Bubble nucleation begins when the nucleation rate per Hubble volume equals the Hubble rate $H(t)$, $\Gamma(t) = H(t)^4$. We choose $t=0$ as the nucleation time and consider the bubble nucleation rate expanded around that time as
\be \label{eq:Gamma}
    \Gamma(t) = H_0^4 \exp\left[\beta t - \frac12 \gamma^2 t^2\right] \,,
\ee
where $H_0 = H(t=0)$. The fraction of the Universe that is trapped in the false vacuum state evolves on average as~\cite{Guth:1982pn}
\be \label{eq:Fbar}
	\bar{F}(t) = \exp\left[-\frac{4\pi}{3} \int_{-\infty}^t \! \td t_n\, \Gamma(t_n) a(t_n)^3 R(t,t_n)^3\right] \,,
\ee
where $R(t_n,t)$ is the comoving radius of a bubble nucleated at time $t_n$ and $a(t)$ is the scale factor of the universe. We assume that the wall velocity is $v_w=1$, as appropriate for strong transitions~\cite{Lewicki:2021pgr, Laurent:2022jrs}.

The potential energy $\Delta V$ associated with the false vacuum state provides vacuum energy density $\rho_v$. If it becomes a dominant contribution of the total energy density before the transition starts, it induces a period of thermal inflation. In the phase transition, the nucleation and expansion of bubbles converts the vacuum energy into radiation, ending the inflation period. The expansion rate of the universe throughout the transition is described by the Friedmann equations
\bea\label{eq:Friedmann}
    &3 M_P^2 H^2 = \rho_v + \rho_r \,, \\
    &\frac{\td \rho_r}{\td t} +4 H \rho_r = - \frac{\td \rho_v}{\td t} \,,
\eea
where $\rho_v=\Delta V \bar{F}(t)$ and $\rho_r$ denotes the radiation energy density, and we have used the fact that the energy in relativistic bubble walls redshifts as radiation~\cite{Lewicki:2023ioy}. 

The thermal inflation ends at $t\approx t_{\rm max}$ when $\bar{F}(t_{\rm max}) = 0.3$. The temperature right after that time can be approximated as
\be \label{eq:Treh}
    T_{\rm reh} \approx 8.5\times 10^8\,{\rm GeV} \left[\frac{g_*}{100}\right]^{-1/4} \sqrt{\frac{H_0}{\rm GeV}} \,,
\ee
where $g_*$ denotes the effective number of relativistic energy degrees of freedom at $T=T_{\rm reh}$. The largest comoving wavenumber that exits horizon during the thermal inflation is $k_{\rm max} \approx a(t_{\rm max}) H(t_{\rm max})$, which can be approximated as
\be \label{eq:kmax}
    k_{\rm max} \approx 1.6\times 10^{-7} {\rm Hz} \,\bigg[ \frac{g_*}{100} \bigg]^{\frac12} \bigg[ \frac{100}{g_{*s}} \bigg]^{\frac13} \frac{T_{\rm reh}}{\rm GeV} \,,
\ee
where $g_{*s}$ denotes the effective numbers of relativistic entropy degrees of freedom at $T=T_{\rm reh}$.

\begin{figure}
\centering
\includegraphics[width=0.46\textwidth]{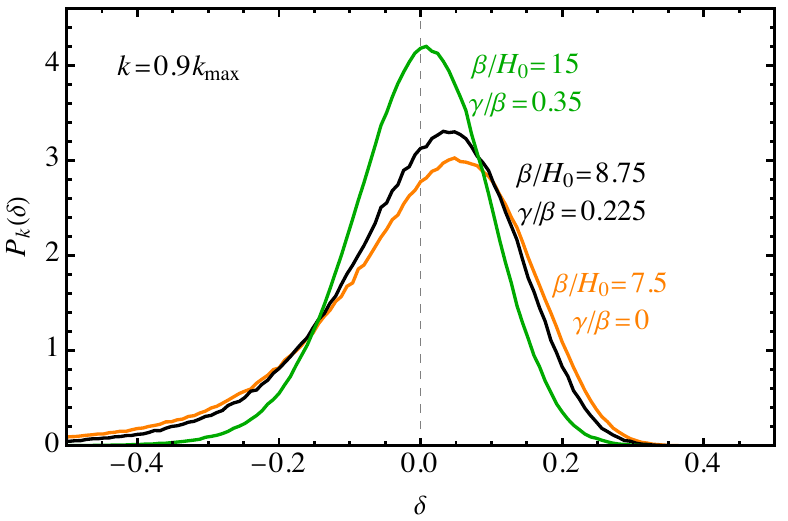}
\caption{The distribution of $\delta$ at $k=0.9k_{\rm max}$ for benchmark points both giving roughly the observed PBH abundance in the asteroidal mass window if $T_{\rm reh} = 10^6\,$GeV.}
\label{fig:Pkdelta}
\end{figure}

We compute false vacuum fraction $F_k(t)$ in different patches $k$ using the \href{https://github.com/vianvask/deltaPT}{deltaPT} code. We generate $10^6$ realizations of $F_k(t)$ and compute the distribution $P_k(\delta)$ of density contrast $\delta = \rho_k(t_k)/\bar{\rho}(t_k) - 1$ at the time $t_k$ when the scale $k$ re-enters horizon, $a(t_k) H(t_k) = k$. The resulting $P_k(\delta)$ at $k=0.9k_{\rm max}$ is shown for three benchmark points in Fig.~\ref{fig:Pkdelta}. Notice that the distribution for the benchmark point with $\gamma = 0$ has a strong negative non-Gaussianity while for larger values of $\gamma$ the distribution becomes almost Gaussian.

\vspace{5pt}\noindent\textbf{Primordial black holes --} Large density perturbations lead to formation of PBHs as the radiation pressure is not enough to prevent the collapse of the overdense patch when it re-enters horizon~\cite{Carr:1974nx,Carr:1975qj}. For any given scale $k$ the fraction of the total energy density that collapses to PBHs of mass $M$ is given by~\cite{Carr:1975qj,Gow:2020bzo,Lewicki:2024ghw}
\bea
    \beta_k(M) &= \int_{\delta_c} \td \delta \frac{M}{M_k} P_k(\delta) \,\delta_D\!\left(\ln \frac{M}{M(\delta)}\right) \\
    &= \frac{\kappa}{\gamma} \left(\frac{M}{\kappa M_k}\right)^{\!1+\frac{1}{\gamma}} P_k(\delta(M)) \,,
\eea
where $M_k = M_p^2/(2H(t_k))$ denotes the horizon mass when the scale $k$ re-enters horizon, $M(\delta) = \kappa M_k (\delta - \delta_c)^\gamma$ following the critical scaling law~\cite{Choptuik:1992jv,Niemeyer:1997mt,Niemeyer:1999ak}, $\delta_D$ denotes the Dirac delta function and $\delta(M)$ is the inverse of $M(\delta)$. The parameters $\gamma$, $\kappa$ and $\delta_c$ depend on the sphericity and the profile of the overdensity as well as the equation of state of the Universe~\cite{Musco:2018rwt,Young:2019yug,Musco:2020jjb,Yoo:2020lmg,Franciolini:2022tfm,Musco:2023dak}. We use the fixed values $\gamma = 0.38$, $\kappa = 4.2$ and $\delta_c = 0.55$, corresponding to the perfect radiation-fluid case~\cite{Franciolini:2022tfm}. From $\beta_k(M)$ we obtain the present PBH mass function as~\cite{Lewicki:2024ghw}
\bea
    &\psi(M) = \int \!\td \ln k \, \beta_k(M) \frac{\rho_r(T_k)}{\rho_c} \frac{s(T_0)}{s(T_k)} \\
    &\approx \frac{2.0\times 10^8}{h^2} \frac{g_*}{g_{*s}} \frac{T_{\rm reh}}{\rm GeV} \!\int \!\td \ln k\, \beta_k(M)  \bigg[\frac{H(t_k)}{H_0}\bigg]^{\frac12} ,
\eea
where $\rho_c$ denotes the critical energy density of the Universe, $s(T)$ the entropy density and $T_0$ the present CMB temperature. The total fraction of DM in PBHs is $f_{\rm PBH} = \int \td \ln M \,\psi(M)/\Omega_{\rm DM}$.

We note that the above procedure relies on the assumption that the Universe is homogeneous and radiation dominated at the moment of horizon re-entry of the curvature fluctuations. While this is clearly not the case during the phase transition, we have checked that these conditions are approximately satisfied already at the time when scales $k=0.9k_{\rm max}$ become subhorizon~\cite{Lewicki:2024ghw}. This also applies to the consideration of the generation of the scalar induced component of the GW spectrum.

\begin{figure}
\centering
\includegraphics[width=0.48\textwidth]{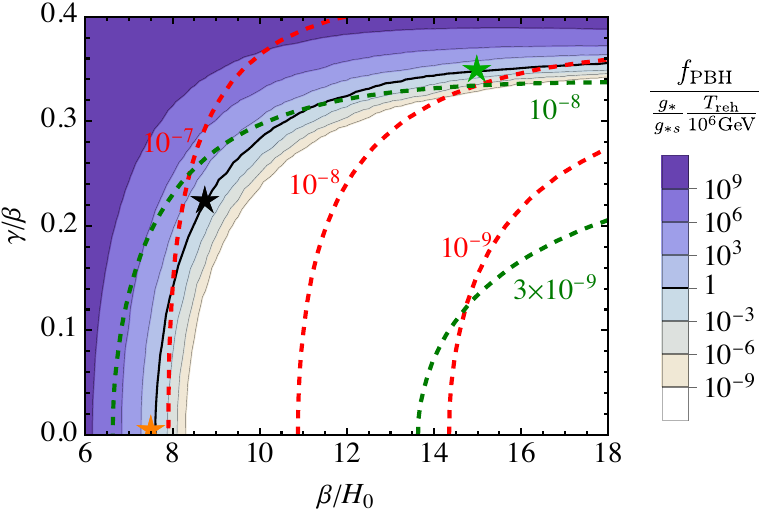}
\caption{The color coding shows the PBH abundance generated in the phase transition and the thick solid contour highlights the contour for which $f_{\rm PBH} = 1$ if $T_{\rm reh} = 10^6\,$GeV. The dashed green and red contours show, respectively, the peak amplitudes of the primary and the secondary contributions to the present GW spectrum. The stars mark the points shown in Figs.~\ref{fig:Pkdelta} and \ref{fig:OmegaGW}.}
\label{fig:fpbh}
\end{figure}

We show the total PBH abundance by the color coding in Fig.~\ref{fig:fpbh}. For small values of $\gamma$, a large PBH abundance is obtained for relatively low values of $\beta/H_0 \lesssim 8$. As $\beta$ increases, the transition becomes faster, and the probability of generating large perturbations decreases. This can be altered by large $\gamma$ values for which the nucleation rate is rapidly damped. Large $\gamma$ values effectively prolong the transition so that a large PBH abundance is reachable for transitions with large $\beta$ values if $\gamma/\beta \gtrsim 0.3$.

\vspace{5pt}\noindent\textbf{Gravitational waves --} The total present abundance of the GWs produced in the phase transition is a sum of the primary component $\Omega_{\rm PGW}$ sourced by the collisions of the bubble walls and fluid shells~\cite{Kosowsky:1992vn,Huber:2008hg,Weir:2016tov,Jinno:2016vai,Jinno:2017fby,Konstandin:2017sat,Cutting:2018tjt,Lewicki:2020jiv,Lewicki:2020azd,Cutting:2020nla,Lewicki:2022pdb} and the secondary component $\Omega_{\rm SGW}$ induced by large curvature perturbations~\cite{Tomita:1975kj,Matarrese:1993zf,Mollerach:2003nq,Ananda:2006af,Baumann:2007zm,Acquaviva:2002ud,Domenech:2021ztg}
\be
    \Omega_{\rm GW} h^2 \approx 1.6\times 10^{-5} \!\left[ \frac{g_*}{100} \right] \!\left[ \frac{g_{*s}}{100} \right]^{\!-\frac43} \left[ \Omega_{\rm PGW} + \Omega_{\rm SGW} \right] \,.
\ee


For the primary component, we use the results derived in~\cite{Lewicki:2020azd,Lewicki:2022pdb}. The computations in~\cite{Lewicki:2020azd,Lewicki:2022pdb} considered the first-order expansion of the nucleation rate, $\Gamma \propto \exp(\tilde\beta t)$. However, Ref.~\cite{Jinno:2017ixd} found that the second order term does not significantly change the shape of the spectrum and we expect that the peak amplitude and frequency are given by the mean bubble separation~\cite{Turner:1992tz,Enqvist:1991xw}
\be
    R_p = \left[ \int_{-\infty}^{t_p} \td t_n \frac{a(t_n)^3}{a(t_p)^3} \bar{F}(t_n) \Gamma(t_n) \right]^{-\frac13} \,,
\ee
where $t_p$ denotes the percolation time, $\bar{F}(t_p) = 1/e$, and the dependence on the nucleation rate enters only through that quantity. So, we compute $R_p$ with the nucleation rate~\eqref{eq:Gamma} and use it in the results of~\cite{Lewicki:2020azd,Lewicki:2022pdb} once we also compute the relation $\tilde{\beta}(R_p)$ for the nucleation rate $\Gamma \propto \exp(\tilde\beta t)$. The primary GW spectrum is a broken power-law
\bea \label{eq:Omega_PT}
    \Omega_{\rm PGW} = \left[ \frac{\tilde{\beta}(R_p)}{H_p} \right]^{-2} \frac{A (a+b)^c S_H(k,k_{\rm max})}{\left[b \left({k}/{k_p}\right)^{\!-a/c} + a \left({k}/{k_p}\right)^{\!b/c}\right]^{\!c}} \,,
\eea
where $k_p \approx 0.7 k_{\rm max} \tilde{\beta}(R_p)/H_p$, $H_p = H(t_p)$,  $A=5.1\times 10^{-2}$, $a=b=2.4$, $c=4$~\cite{Lewicki:2022pdb} and $S_H(k,k_{\rm max})$ accounts for the causality-limited tail of the spectrum at $k\lesssim k_{\rm max}$~\cite{Caprini:2009fx,Franciolini:2023wjm} that we approximate as in~\cite{Ellis:2023oxs}.

The abundance of the secondary GWs is dominantly determined by the characteristic amplitude of perturbations described by the curvature power spectrum $\mathcal{P}_\zeta$. Neglecting the corrections arising from the non-Gaussianities~\cite{Unal:2018yaa,Cai:2018dig,Yuan:2020iwf,Adshead:2021hnm,Abe:2022xur,Ellis:2023oxs}, the spectrum of the secondary GWs is given by~\cite{Kohri:2018awv,Espinosa:2018eve,Inomata:2019yww}
\bea
    \Omega_{\rm SGW} \approx \frac{1}{3} \int_1^\infty \!\!\td t & \int_{0}^{1} \!\td s \,\mathcal{I}_{t,s}^2  \left [ \frac{(t^2-1)(1-s^2)}{t^2-s^2} \right ]^2 \\
    &\times \mathcal{P}_\zeta\left(k\frac{t-s}{2}\right) \mathcal{P}_\zeta\left(k\frac{t+s}{2}\right) ,
\eea
where $\mathcal{I}_{t,s}$ denotes the transfer function given e.g. in~\cite{Inomata:2019yww}. We compute $\mathcal{P}_\zeta$ from the variance of the density perturbations using the linear relation $\zeta = 9 \delta/4$ between the density contrast and the curvature perturbation at the time of horizon crossing, which holds under the assumption of radiation dominance.

The peak amplitudes of the primary and the secondary GWs are shown in Fig.~\ref{fig:fpbh}. We stress the difference between the shape of peak amplitude and PBH abundance contours arises due to the much larger impact of non-Gaussianities on PBH formation than on secondary GWs. Transitions with smaller value of $\gamma$ produce strong negative non-Gaussianity (see Fig.~\ref{fig:Pkdelta}), which reduces the probability of PBH formation while leaving the secondary GW signals roughly unaffected. For the same reason, the primary GWs dominate over the secondary GWs in the parameter space relevant for PBH formation if $\gamma$ and $\beta$ are sufficiently large, while for small $\gamma$ and $\beta$ values the amplitude of the secondary GWs is higher.

\begin{figure}
\centering
\includegraphics[width=0.46\textwidth]{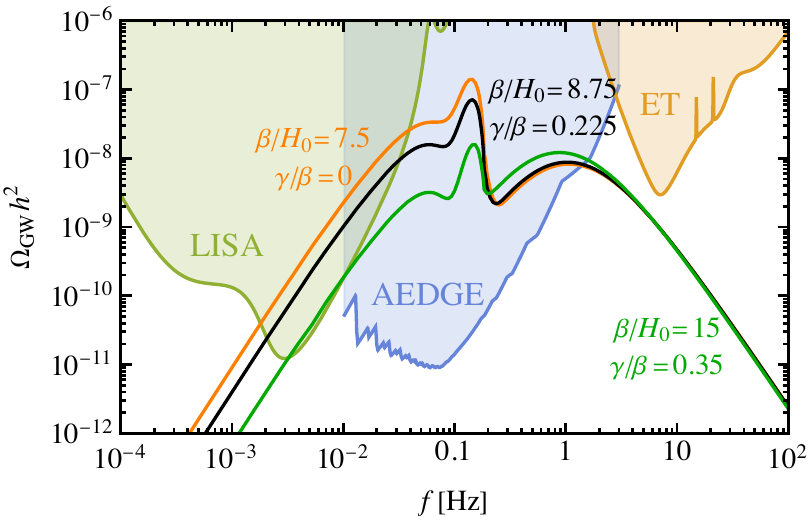}
\caption{The GW spectrum for the same benchmark points as in Fig.~\ref{fig:Pkdelta} with $T_{\rm reh} = 10^6\,$GeV. The shaded regions in the right panel indicate the instantaneous sky averaged sensitivities of LISA, AEDGE and ET.}
\label{fig:OmegaGW}
\end{figure}

The GW spectra resulting from the benchmark points shown in Fig.~\ref{fig:Pkdelta} are shown in Fig.~\ref{fig:OmegaGW}. This plot illustrates the differences in the GW spectra that correspond to the same PBH abundance in the asteroidal mass window. These spectra would be observed with a large signal-to-noise ratio (SNR) by
ET~\cite{Punturo:2010zz, Hild:2010id}, AEDGE~\cite{AEDGE:2019nxb,Badurina:2021rgt} and LISA~\cite{Bartolo:2016ami, Caprini:2019pxz, LISACosmologyWorkingGroup:2022jok} experiments, with AEDGE seeing the peak frequencies. The spectra are characterized by two distinct peaks: The peak at larger frequency originates from the primary GWs and corresponds to the mean bubble separation $R_p$. The lower frequency peak is at the scale of the horizon size at the end of the thermal inflation, $k=k_{\rm max}$ and originates from the secondary GWs. The main difference between the spectra is in the secondary contribution and is caused by the non-Gaussianities (see Fig.~\ref{fig:Pkdelta}) that affect the PBH formation.

\vspace{5pt}\noindent\textbf{Realistic particle physics example --} Slow and strongly supercooled phase transitions can be realized in classically scale invariant models. For illustration, we consider a simple potential 
\be \label{eq:SIpotential}
    V(\phi, T) = V_0 + \frac{3g^4}{4\pi^2} \phi^4 \left[ \ln \frac{\phi^2}{v^2}-\frac{1}{2} \right] + \frac{g^2 T^2}{2} \phi^2\, ,
\ee
where $T$ denotes the radiation bath temperature, $v$ is the $T=0$ vacuum expectation value of $\phi$ and $V_0$ is a constant chosen so that $V(T=0,\phi=v) = 0$. This is the one-loop high-$T$  effective potential in classically scale invariant scalar electrodynamics where $g$ is the U$(1)$ gauge coupling and provides a reasonable approximation in various particle physics models~\cite{Jinno:2016knw, Iso:2017uuu, Marzola:2017jzl, Salvio:2019wcp, Kierkla:2022odc, Kierkla:2023von, Prokopec:2018tnq, Marzo:2018nov, Baratella:2018pxi, VonHarling:2019rgb, Aoki:2019mlt, DelleRose:2019pgi, Wang:2020jrd, Ellis:2019oqb, Ellis:2020nnr, Baldes:2020kam, Baldes:2021aph, Lewicki:2021xku, Gouttenoire:2023pxh}. For $g\lesssim 0.5$ the transition is so strongly supercooled that the vacuum energy of the false vacuum causes a period of thermal inflation before the transition to the true vacuum at $\phi \sim v$ occurs~\cite{Lewicki:2020azd}.

\begin{figure}
\centering
\includegraphics[width=0.46\textwidth]{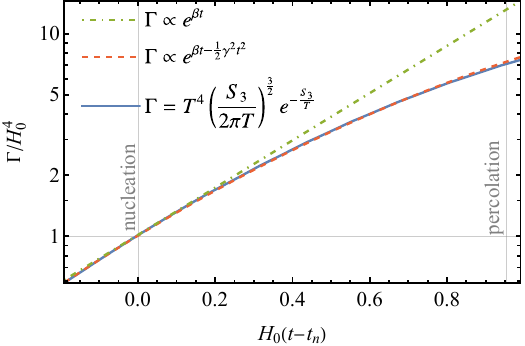}
\caption{Accuracy of the expansion of the decay rate for $v=3\times10^7$GeV and $g=0.33$ that gives $R_p H_p \approx 1$. The vertical lines show the nucleation time (left) and the percolation time (right).}
\label{fig:DecayExpansion}
\end{figure}

The decay rate of the false vacuum due to thermal fluctuations reads~\cite{Linde:1980tt,Linde:1981zj}:
\be \label{eq:gamma:2.1}
    \Gamma= \left(\frac{S_3}{2\pi T}\right)^{\frac{3}{2}} T^4 e^{-S_3/T} \,,
\ee
where $S_3$ is the action for the solution corresponding to the nucleating bubble. We expand the nucleation rate as in~\eqref{eq:Gamma}. It is convenient to rewrite the decay rate using a dimensionful constant $\Lambda$ as $\Gamma=\Lambda^4 e^{-S(T)}$ defining
\be
    S(T) = \frac{S_3}{T} - \ln\left[\left(\frac{T}{\Lambda}\right)^4 \left( \frac{1}{2\pi} \frac{S_3(T)}{T} \right)^{\frac32} \right] \,.
\ee
The first-order term is the commonly used inverse duration of the phase transition (in Hubble units),
\be
    \frac{\beta}{H} = T\frac{\td S(T)}{\td T} \, ,
\ee
where we used the relation $dT/dt=-HT$. In very fast transitions with $\beta/H \gg 1$ it is typical to cut the expansion at the first term. However, for slow transitions, this is not a good approximation~\cite{Megevand:2016lpr, Jinno:2017ixd, Ellis:2018mja, Cutting:2018tjt, Levi:2022bzt} and higher order terms are needed. The coefficient of the second order term reads
\be
    \left( \frac{\gamma}{H} \right)^2 = \frac{\beta}{H}\left(1+ \frac{T}{H}\frac{\td H}{\td T}\right) + T^2\frac{\td^2 S(T)}{\td T^2} \, ,
\ee
and it is straightforward to continue the series. The progress of the transition and evolution of the Hubble rate is described by Eqs.~\eqref{eq:Fbar} and~\eqref{eq:Friedmann} which we solve numerically for each set of model parameters. 

To ensure that the phase transition completes, one has to check that the physical volume of the false vacuum, $V_{\rm{false}} \propto a(t)^3 \bar{F}(t)$, decreases~\cite{Turner:1992tz,Ellis:2018mja}, $\td V_{\rm{false}}/\td t < 0$, which is commonly done at the percolation time $t_p$. For all the parameter space we considered, this coincides with the simple criterion $R_p H_p < 1$. 

We illustrate the accuracy of the second order expansion of the action in Fig.~\ref{fig:DecayExpansion} in the range of relevant temperatures between the nucleation and percolation for a point with $v=3 \times 10^7$GeV and $g=0.33$ that predicts $R_p H_p \approx 1$. In such case near the percolation limit the expansion of the action is the most problematic as it needs to reproduce the rate in a large time range. However, even there we find that, for all the parameter space we consider, the expansion up to the second order is a valid approximation and, while we mostly use the full unexpanded action, all further results change negligibly when using the expansion instead. We checked that the relevant parameters change by at most a few percent when switching from the full decay rate to its second-order expansion.

\begin{figure}
\centering
\includegraphics[width=0.49\textwidth]{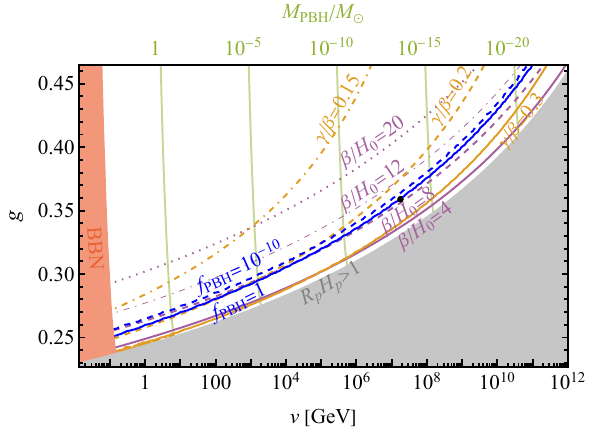}
\caption{The purple and orange contours show the parameters $\beta/H_0$ and $\gamma/\beta$ defining the expansion of the nucleation rate up to second order (see Eq.~\eqref{eq:Gamma}) as a function of model parameters $v$ and $g$. The blue contours showing the PBH abundance are normalized to the observed DM abundance and the green contours show the mean PBH masses. The red region is excluded by BBN constraint on the reheating temperature after the transition and the grey region by the percolation criterion that matches the condition $R_p H_p < 1$.}
\label{fig:ExpansionParameterSpace}
\end{figure}

In Fig.~\ref{fig:ExpansionParameterSpace} we show the values of the first and second order coefficients $\beta$ normalized to the Hubble rate (purple) and $\gamma$ normalized to $\beta$ (orange). The latter plateaus quickly for weaker transitions and the value of $\gamma/\beta = 0.1$ is not reached anywhere in the parameter space of interest. The gray area is excluded by the lack of percolation which coincides with the criterion $R_p H_p > 1$ and the red region by the BBN constraint $T_{\rm reh} > 5$\,MeV~\cite{Allahverdi:2020uax}.\footnote{We assume the energy released by the transition instantaneously reheats the SM thermal bath. In more realistic models, hiding from collider experiments typically induces additional constraints for example if the transition occurs in a dark sector~\cite{Elor:2023xbz}.} As seen from Fig.~\ref{fig:fpbh}, taking the second order term into account in the computation of the PBH abundance and the abundance of secondary GWs is important for the values of $\gamma$ found in this model. The blue and green curves in Fig.~\ref{fig:ExpansionParameterSpace} indicate the PBH abundance and mass. The black dot corresponds to the benchmark case shown in black in Figs.~\ref{fig:Pkdelta} and~\ref{fig:OmegaGW}.

\begin{figure}
\centering
\includegraphics[width=0.46\textwidth]{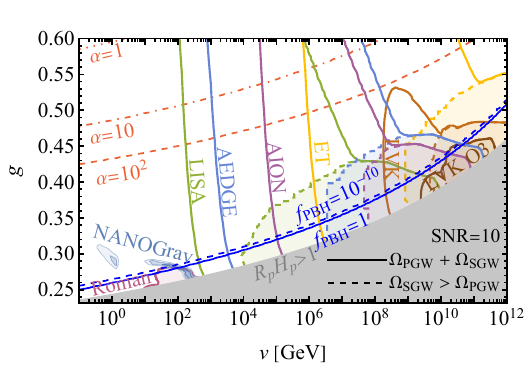}
\caption{Reach of GW experiments: The solid curves indicate the parts of the parameter space that give SNR$>10$ while in the shaded regions the SNR is dominated by the secondary GWs. The light blue area on the left shows the fit to NANOGrav data. The blue contours show the PBH abundance normalized to the observed DM abundance, while the grey region is excluded by the percolation criterion that matches the condition $R_p H_p < 1$.}
\label{fig:GWsParameterSpace}
\end{figure}

In Fig.~\ref{fig:GWsParameterSpace} we show the regions of the model parameter space that are within reach of planned GW experiments with ${\rm SNR} \geq 10$. The solid curves show the detection prospects for the sum of primary and secondary GW spectra while the dashed lines show the region where the secondary GWs dominate the SNR. We include the future reach of LVK using the design sensitivity of LIGO~\cite{LIGOScientific:2014pky, LIGOScientific:2016fpe, LIGOScientific:2019vic}, LISA~\cite{Bartolo:2016ami, Caprini:2019pxz, LISACosmologyWorkingGroup:2022jok}, AEDGE~\cite{AEDGE:2019nxb,Badurina:2021rgt}, AION~\cite{Badurina:2019hst, Badurina:2021rgt}, ET~\cite{Punturo:2010zz, Hild:2010id} and the Nancy Roman telescope~\cite{Wang:2022sxn}. We also show the exclusion from the current LVK (O3) observations~\cite{KAGRA:2021kbb}, which mostly falls into the region already excluded by PBH overproduction, and the region where the GW background fits the NANOGrav observations~\cite{NANOGrav:2023gor}. The red contours show the corresponding values of the strength of the transition $\alpha = \Delta V/\rho_r(T_p)$ where $T_p$ is the percolation temperature. We see that, in the entire parameter space of interest, our assumption that the transition is very strong, $\alpha\gg 1$, is satisfied.

\vspace{5pt}\noindent\textbf{Conclusions --} In this paper we have studied the production of PBHs and GWs in slow and strongly supercooled first-order phase transitions. We have shown that including the second-order correction in the expansion of the bubble nucleation rate is necessary and sufficient for accurate predictions. To this end, we have evaluated the full predictions of a realistic model featuring such transitions and compared the results with those from the simplified modelling involving an expansion of the nucleation rate. The relevant parameters obtained by second-order expansion and the full decay rate deviate by at most a few percent.

We have quantified the impact of the nucleation history on the GW signals and PBH production. The GW spectrum contains two peaks corresponding to the horizon size and the characteristic bubble size. We have found that the second-order term affects the relative heights of these peaks. Moreover, we have found that models predicting the same PBH abundance can result in different GW spectra. While the GW amplitude is determined by the typical size of the fluctuations, the PBH abundance is sensitive also to their non-Gaussianity which decreases for larger values of the second-order term.

We have also systematically included the impact of the bubble nucleation on the expansion of the Universe. We have found that, due to this improvement, the well-known percolation criterion coincides with a very simple and intuitive requirement that the size of the bubbles at the time of collision has to be smaller than the size of the horizon, $RH<1$. 

\vspace{5pt}\noindent{ \it Note added} - Since the initial publication of our preprint the importance of the issue of gauge fixing was pointed out in~\cite{Franciolini:2025ztf}. The mismatch between the flat gauge used in our calculation and the comoving gauge used in the computation of the threshold for PBH collapse can have a significant impact on the final production of black holes and gravitational waves. We will address this issue in a future publication.

\vspace{5pt}\noindent\textbf{Acknowledgments --} The work of M.L. and P.T. was supported by the Polish National Agency for Academic Exchange within the Polish Returns Programme under agreement PPN/PPO/2020/1/00013/U/00001 and the Polish National Science Center grant 2023/50/E/ST2/00177. The work of P.T. was supported by the Polish National Science Center grant 2024/53/N/ST2/04009. The work of V.V. was supported by the European Union's Horizon Europe research and innovation program under the Marie Sk\l{}odowska-Curie grant agreement No. 101065736, and by the Estonian Research Council grants PRG803, RVTT3 and RVTT7 and the Center of Excellence program TK202.

\bibliographystyle{elsarticle-num}
\bibliography{main_old}

\end{document}